\documentstyle[12pt,aasms4]{article}

\received{}
\accepted{}
\journalid{}{}
\articleid{}{}

\slugcomment{to be published in ApJ Letters}

\lefthead{Martin et al.}
\righthead{A Companion to Gl~569~B}

\begin{document}


\title{The Discovery of a Companion to the Very Cool Dwarf Gl~569~B  
with the Keck Adaptive Optics Facility}

\author{E. L. Mart\'\i n, C. D. Koresko, S. R. Kulkarni, B. F. Lane}
\affil{Division of Geological and Planetary Sciences, 
California Institute of Technology, MS 150-21, 
Pasadena, CA 91125}

\and 

\author{P. L. Wizinowich}
\affil{W. M. Keck Observatory, 65-1120 Mamalahoa Highway, Kamuela, Hi 96743}

\centerline{contact e-mail address: ege@gps.caltech.edu} 

\begin{abstract}

We report observations obtained with the Keck adaptive optics facility 
of the nearby (d=9.8~pc) binary Gl~569. The system was known to be composed 
of a cool primary (dM2) and a very cool secondary (dM8.5) with a separation of 5" 
(49 Astronomical Units). 
We have found that Gl~569~B is itself double with a separation of 
only 0".101$\pm$0".002 (1 Astronomical Unit).  
This detection demonstrates the superb spatial resolution that can 
be achieved with adaptive optics at Keck.  
The difference in brightness between Gl~569~B and the companion is $\sim$0.5 magnitudes  
in the J, H and K' bands. Thus, both objects have similarly red colors  
and very likely constitute a very low-mass binary system. 
For reasonable assumptions about the age (0.12~Gyr--1.0~Gyr) 
and total mass of the system (0.09~M$_\odot$--0.15~M$_\odot$), 
we estimate that the orbital period is $\sim$3 years. 
Follow-up observations will allow us to obtain an astrometric 
orbit solution and will yield direct dynamical masses that can constrain 
evolutionary models of very low-mass stars and brown dwarfs.

\end{abstract}


\section{Introduction}

Substellar objects do not have enough mass to settle on the  
hydrogen-burning main sequence, and therefore their luminosity and surface 
temperature are very age dependent. The substellar mass limit has 
been determined with theoretical models. For solar composition, 
Kumar (1963) obtained a limiting mass of 0.07~M$_\odot$. Recent 
determinations using non-gray model atmospheres reached a similar 
result (0.075~M$_\odot$; Baraffe et al. 1998). The limiting mass 
is larger for low metallicity. 

Lithium is a fragile element in stellar interiors that gets destroyed 
at temperatures below those necessary for hydrogen fusion. 
Very low-mass (VLM) stars and brown dwarfs (BDs) with masses above  
$\sim$0.060~M$_\odot$ destroy their initial lithium reservoir while they are fully convective. 
BDs with masses below $\sim$0.06~M$_\odot$ never burn lithium 
(Magazz\`u, Mart\'\i n \& Rebolo 1993; Nelson, Rappaport \& Chiang 1993). 
Spectroscopic observations of BD candidates can detect lithium if this element has 
been preserved, and provide information about the mass of the object. 
This is the so-called lithium test for BDs, first proposed by Rebolo, Mart\'\i n \& Magazz\`u (1992).  
Although early lithium searches in low-luminosity dwarfs were unsuccessful  
(Mart\'\i n, Rebolo \& Magazz\`u 1994; Marcy, Basri \& Graham 1994), the test 
gave positive results in the Pleiades open cluster, confirming the existence 
of free-floating BDs  
(Rebolo et al. 1996).  
The combination of lithium abundances, effective temperatures and luminosities with 
evolutionary models provide the only means of estimating 
ages and masses for single VLM stars and BDs (Tinney 1998; 
Mart\'\i n, Basri \& Zapatero Osorio 1999). However, these masses rely on theoretical 
models that have not been well tested. Perhaps the strongest test to the models comes 
from binary systems where masses can be measured from the orbital parameters. 
The most accurate mass determinations are derived  from   
eclipsing binaries. The problem is that they are rare. No VLM stars or BDs are known 
to belong to an eclipsing binary system. 

Wide binaries also yield dynamical masses. With the advent of high spatial resolution 
instrumentation (Hubble Space Telescope, hereafter HST; adaptive optics, 
hereafter AO; speckle techniques; interferometry) it is possible to 
resolve binaries with smaller separations and consequently shorter periods. 
Leinert et al. (1994) discovered with speckles a BD candidate companion 
to the dM5.5 nearby star LHS~1070. The spectral type of the companion 
is dM8 (Leinert et al. 1997) and the orbital period is about 20 years. 
Mart\'\i n, Brandner \& Basri (1999) found with HST the first resolved binary system 
(separation = 0".275 or 5 AU) 
where both components are L-type BDs. They estimated that the 
 orbital period could be about 30 years. 
Two more L-type binaries with slightly larger separations 
have recently been found by Koerner et al. (1999) 
using Keck near-infrared imaging. Two BDs companions are also 
known with separations of several arcseconds 
from early-M stars (Nakajima et al. 1995; Rebolo et al. 1998).  

Gl~569 is a nearby star (d=9.8~pc) with high levels of chromospheric and 
coronal activity. Forrest, Skrutskie \& Shure (1988) reported a possible 
BD companion separated 5" from the primary. 
Henry \& Kirkpatrick (1990) obtained a low resolution spectrum and classified it 
as an M8.5 dwarf. They estimated a mass of 0.09~M$_\odot \pm$0.02~M$_\odot$. 
Magazz\`u et al. (1993) did not detect lithium in Gl~569~B, even though its 
position in the HR diagram suggest that it is a very young object. 
In this paper we show that Gl~569~B is  
a binary, and hence the Gl~569 system is at least triple. 
We estimate that the orbital period of the VLM binary could  
be about 3 years for a circular orbit, or shorter for an elliptical orbit, 
implying that dynamical masses can be 
obtained in a relatively short time.

\section{Observations, Data Analysis and Results}

We observed with the Keck~II Adaptive Optics Facility (KAOF) on August 28th, 
1999. The instrument has been described elsewhere (Wizinowich et al. 1998). 
We used the KCAM camera, which has a plate scale of 0".0175~pix$^{-1}$ 
and a square field of view of 4".5 x 4".5. 
Flatfield correction was made using twilight exposures. 
Sky subtraction was performed using images of a field adjacent 
to Gl~569 observed inmediately after 
the science images. We used the DAOPHOT package available in the IRAF environment for 
data reduction and analysis. 

Using a neutral density (ND2; attenuation of $\sim$100) filter and 
co-adding short exposures, we were 
able to detect the fainter components of the multiple system in the J, H and K' filters  
without saturating the primary. The primary star is not double and serves as a 
PSF reference in the field. 
In Fig.~1 we show a 60~s exposure on KCAM, consisting of 30 co-adds,
in the K'-band. Gl~569~A looks asymmetric because it is 
cut off by the edge of the detector, but it is clearly not double or elongated in 
the same orientation as Gl~569~B. 
The edge of the detector does not correspond to 
the drawn box. 
We have rotated the frame by 251 degrees and transposed it in order to present the 
image of the binary in 
the standard orientation where North is up and East to the left. Using Gl~569~A, 
we estimate Strehl ratios of 9\%, 20\% and 40\% 
in the J, H and K' filters (with ND2 filter), respectively.   
We also obtained 18 images in the H-band (without ND2 filter) of Gl~569~B alone, 
with the primary out of the field of view. The two components of the 
binary system were clearly resolved in all 
the frames. In Fig.~2 we show a typical exposure of 3 seconds. Each contour line  
represents half the intensity of the previous one, and
the lowest is 2\% of the maximum intensity. 

Any ghost from Gl~569~Ba would likely not be of comparable magnitude to 
the object (Gl~569~Bb has about 65\% of the flux of Gl~569~Ba).  Moreover, 
if Gl~569~Bb were a ghost from Gl~569~Ba there would likely also be a nearby ghost for Gl~569~A of
the same relative magnitude, but there isn't.  
That means that any ghost would have to have Gl~569~A as its source. 
This is ruled out by the fact that we moved and rotated the binary on the detector, 
observing it in three different quadrants, 
and the separation and magnitude difference between Gl~569~Ba and b remained the same. 
Furthermore, the contrast in the J-band between Gl~569~A and Gl~569~Ba is larger than 
that in the K-band by 1 magnitude (Table~1) because of the cooler temperature of the latter object. 
If Gl~569~Bb was a ghost due to Gl~569~A, it should have the same color. However, we find that 
Gl~569~Bb has similar color than Gl~569~Ba, indicating that it is a cooler object than 
Gl~569~A.

We obtained relative astrometry and photometry by fitting the point spread function with a gaussian  
function. The results are listed in Table~1. No photometric standard was observed in the same night. 
Therefore, we present differential magnitudes only, which are based on a 150~s J-band
frame, a 30~s H-band frame and a 60~s K'-band frame (all taken with an
ND2 filter in the beam). We also measured the H differential magnitude  
in the dithered exposures of 3~s without ND2 filter, and found the same value as in the image 
with the ND2 filter. 
The J-H and H-K' colors of both components of Gl~569~B are the same within our error bars. 
It is not surprising that VLM dwarfs  
with similar absolute magnitudes have nearly identical colors. 
The likelihood that the companion to Gl~569~B is 
a background star is less than 0.1\% because of the very low-density of late M dwarfs 
and giants within the small volume covered by our images. The galactic latitude of 
Gl~569 is quite high (+59.4 degrees). In fact, no objects other 
than those of the Gl~569 system appear in any of our frames. 

\section{Discussion}

Forrest et al. (1988) presented $IHK$ photometry for Gl~569~B. They noted that the 
object is more luminous than other field dwarfs of similar temperature, and 
attributed it to young age. However, Magazz\`u et al. (1993) failed to detect 
lithium in it, and constrained its age to be older than 0.1~Gyr and the mass 
to be larger than 0.06~M$_\odot$. The fact that this object is a binary solves the 
problem of undertanding why it is overluminous but has depleted lithium. The relatively 
high luminosity is due to its binary nature, with components of similar brightness, 
not to a very young age. Nevertheless, as 
discussed below, the system is probably not old. 

We have combined the absolute photometry provided by Forrest et al. 
($I$=13.88$\pm$0.2; $H$=10.16$\pm$0.1; $K$=9.56$\pm$0.1) with our differential 
magnitudes (Table~1) to place Gl~569~Ba and Bb in a color-magnitude diagram (CMD). 
In Fig.~3, we compare Gl~569~B with other very late M dwarfs in the 
Pleiades and the field, which have infrared photometry available in the literature 
(Festin 1998; Leggett, Allard \& Hauschildt 1999; Zapatero Osorio, Mart\'\i n \& Rebolo 1997). A trend 
of increasing H-K color for fainter M$_H$ can be seen in the CMD, but there is 
considerable dispersion.  

Allard et al. (1997), Hauschildt et al. (1999), and Allard (1999) have 
presented model atmospheres for ultracool dwarfs in the 
temperature range 3500~K to 1500~K. The Nextgen set does not include 
any dust effects, while the Dusty set includes dust formation and settling. 
We have used theoretical isochrones computed using Nextgen and Dusty atmospheres 
(Chabrier \& Baraffe 1999) for comparison with the data in Fig.~3. The Nextgen isochrones 
are too blue and do not fit the fainter and redder objects. The Dusty ones  
give a better fit, but the two components of Gl~569~B do not 
fall onto the same isochrone. The problem could be due to the presence 
of strong steam absorption in the H-band which is not well reproduced by the models 
(e.g. Allard 1999). Another possibility is that Gl~569~Ba is itself 
an unresolved binary. Follow-up high-resolution spectroscopic observations are needed 
to test this hypothesis. 
 
The Gl~569~B system must be older than the Pleiades cluster (age$\sim$0.11~Gyr; 
Mart\'\i n et al. 1998) 
because of the lack of lithium in its optical spectrum. On the other hand, 
the system is probably not very old because Gl~569~A has  
strong radio and X-ray emission (Pallavicini, Tagliaferri \& Stella 1990). The high level of chromospheric 
and coronal activity suggests a young age because the star is apparently single. In particular, 
it is not a tidally locked spectroscopic binary (Marcy \& Benitz 1989).  
The position of Gl~569~Bb in the CMD close to the 
Dusty isochrone for 0.1~Gyr may also indicate youth.   
For the purpose of estimating the orbital period of the system, which is 
of interest for follow-up observations, we assume that the age is in the range 
0.12 and 1~Gyr. We derive for the components of Gl~569~B absolute K-band 
magnitudes of 10.15$\pm$0.12 and 10.60$\pm$0.12 from Forrest et al.'s absolute photometry and our differential photometry. According to the Lyon Dusty models, 
for an age of 0.12~Gyr, the masses that correspond to those absolute K magnitudes are 
0.051~M$_\odot$ and 0.042~M$_\odot$, respectively. For an age of 1~Gyr, the corresponding 
masses are 0.087~M$_\odot$ and 0.062~M$_\odot$. While Gl~569~Ba could be either a VLM star 
or a massive BD, depending on its age, Gl~569~Bb is likely to be substellar. 
A lithium detection in Gl~569~Bb would be very interesting for constraining its age. 

The total mass of the system inferred from the Lyon models 
is 0.15~M$_\odot$ for an age of 1~Gyr. Assuming that 
the observed separation is close to the mean semimajor axis, we estimate an orbital period of 
2.58 years. For an age of 0.12~Gyr, the total mass is 0.09~M$_\odot$, and 
the period is 3.33 years. Basri \& Mart\'\i n 
(1999) have determined that the short-period BD binary PPl~15 has a very eccentric orbit. 
The period of Gl~569~B could be shorter if the orbit has a very large 
eccentricity because 
it is more likely to catch it near apastron than near periastron. 
We plan to reobserve it for detecting the orbital motion. 
The orbital parameters of Gl~569~B will allow us to determine if the components 
have substellar masses. It will also be a first point in the mass-luminosity relationship 
for dwarfs later than M8. This binary system is a powerful test for 
evolutionary models of VLM stars and BDs.

\acknowledgments

{\it Acknowledgments}: 
Data presented herein were
obtained at the W.M. Keck Observatory, which is operated as a scientific
partnership among the California Institute of Technology, the University of
California and the National Aeronautics and Space Administration.  The
Observatory was made possible by the generous financial support of the W.M.
Keck Foundation. 
This research has made use of the Simbad database, operated at CDS, Strasbourg, France. 
We would like to acknowledge James Larkin and Ian McLean for
providing KCAM, Scott Acton and the other members of the Keck
AO team. We used the theoretical models computed by 
France Allard, Isabelle Baraffe, Gilles Chabrier and Peter Hauschildt, and made 
available to us by IB in computer readable format.


\clearpage

\centerline{\bf Figure Captions:}

\figcaption[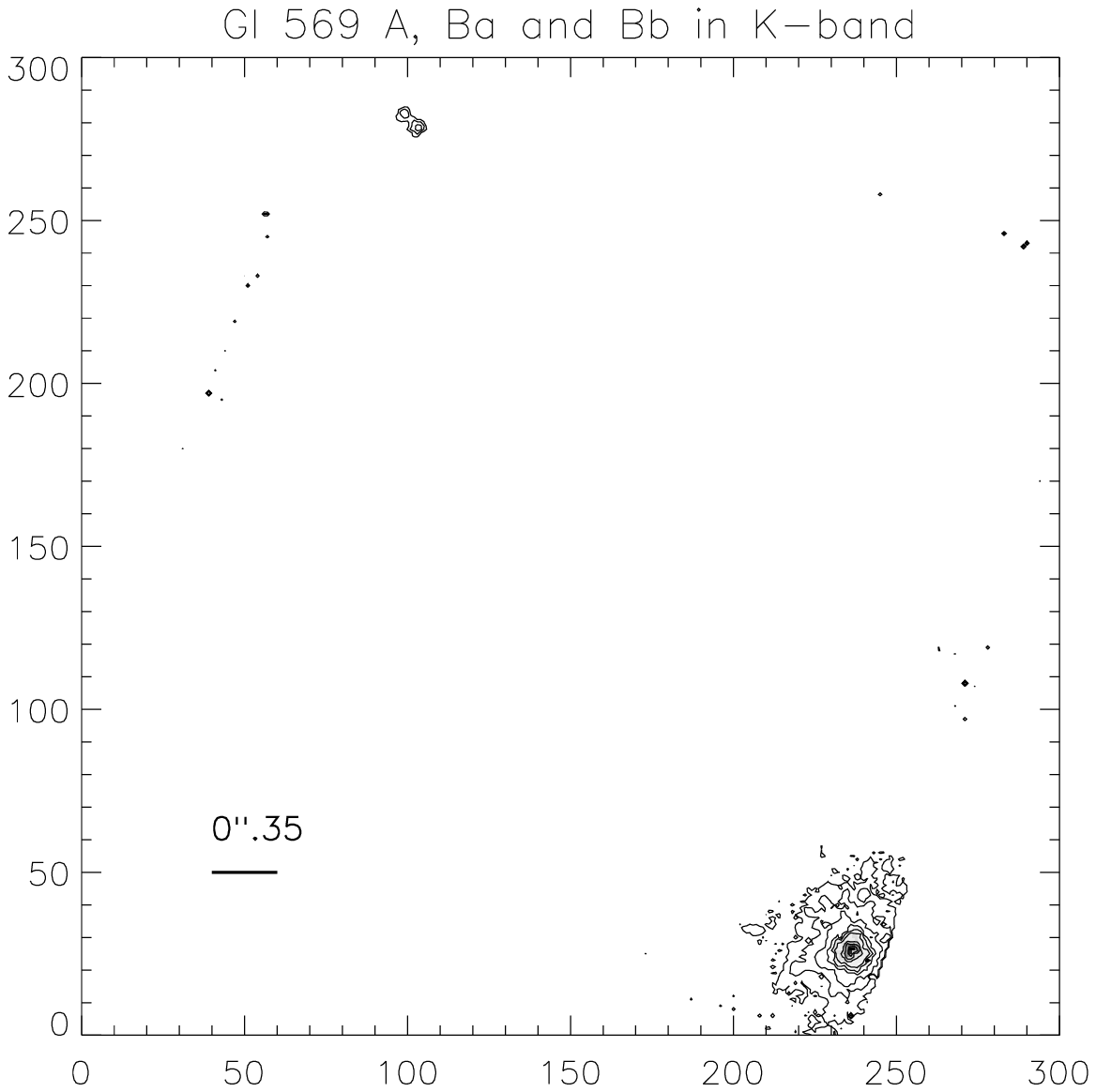]{\label{fig1} A 60~s K' image with neutral 
density filter. The two faint components of the Gl~569~B system are detected 
and the primary, Gl~569~A, is not saturated. 
Contours are peak counts $\times 2^{(9-n)}$. The faintest contour is 0.2\% 
of the maximum peak. The plate scale is 0".0175 per pixel. 
The frame has been rotated so that North is up and East to the left. }

\figcaption[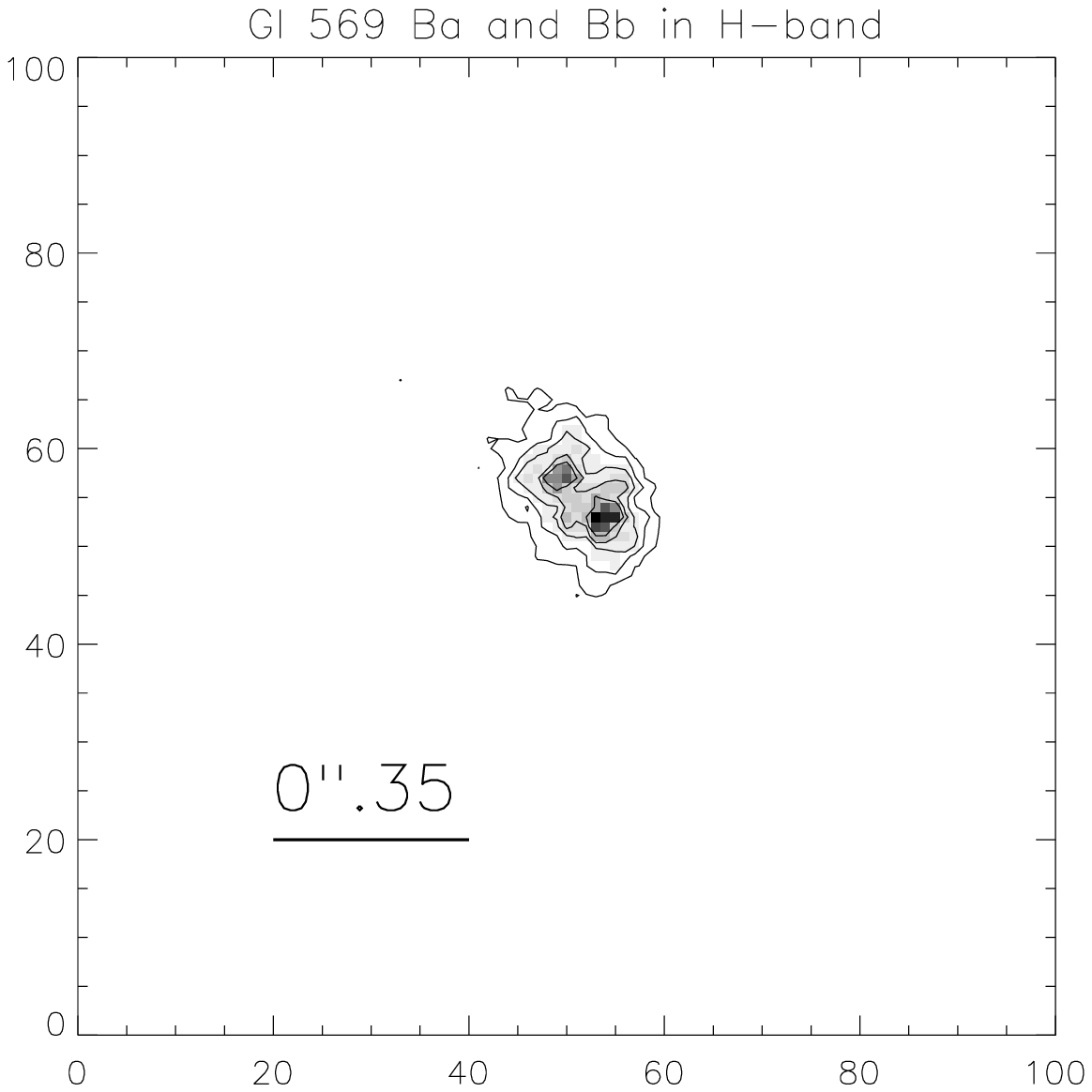]{\label{fig2} A 3~s exposure of the Gl 569 B 
binary in the H-band.  We show a subsection of 
100~pix$^2$ of the full frame. Contours are peak counts $\times 2^{(6-n)}$. 
The faintest contour is 2\% of the maximum peak. 
The orientation and plate scale is the same as 
in the previous figure.}

\figcaption[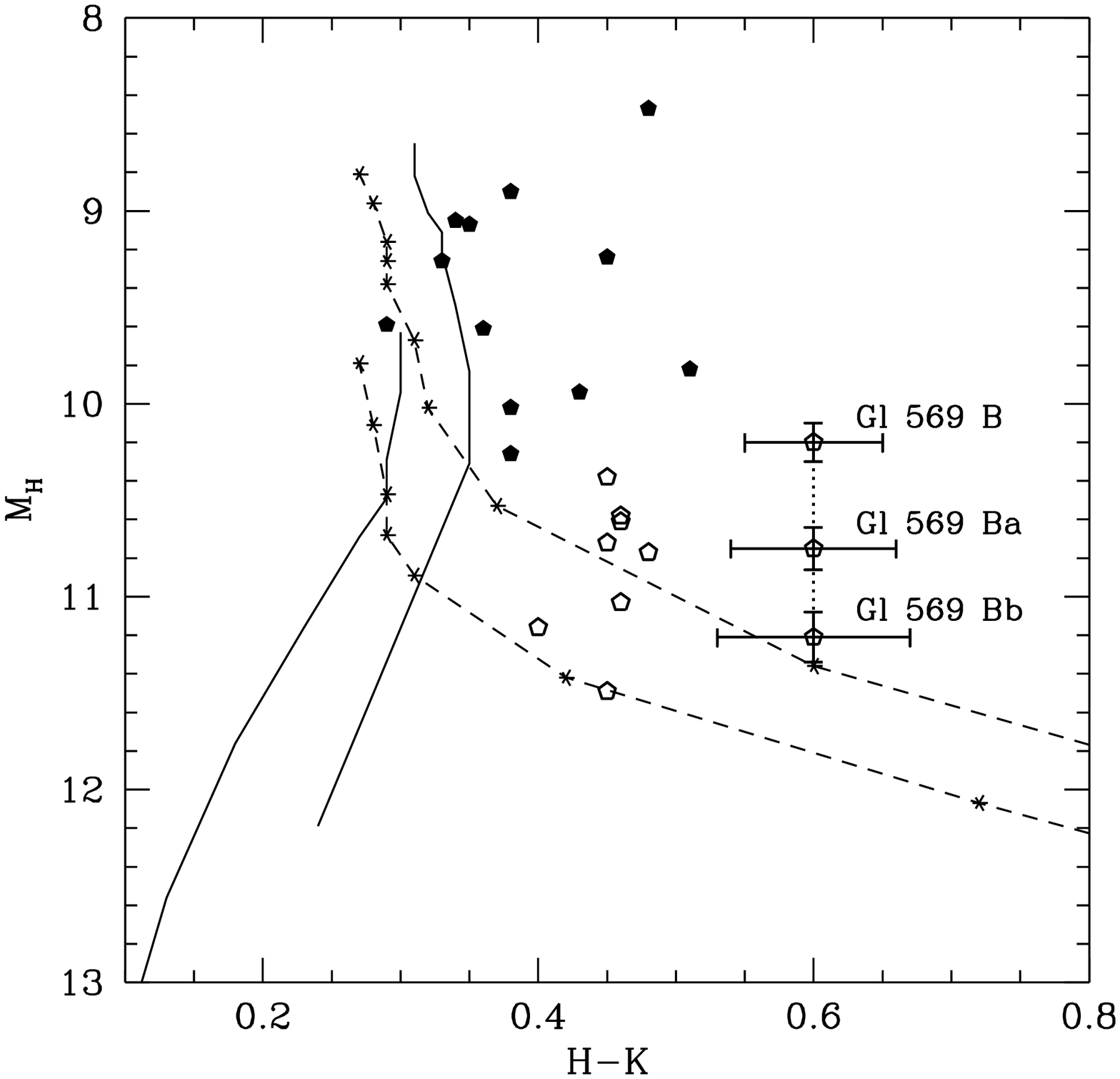]{\label{fig3} An infrared color-magnitude diagram with the Lyon-group 
isochrones (Allard et al. 1997; Baraffe et al. 1998; Hauschildt et al. 1999) 
for ages of 0.1~Gyr and 1~Gyr. The solid lines are for Nextgen model atmospheres and 
the dashed lines are for Dusty models. Masses in solar units of 0.1, 0.09, 0.08, 0.075, 
0.07, 0.06, 0.05, 0.04, 0.03 and 0.02 are marked with six pointed stars on the Dusty isochrones. 
Pleiades BD 
candidates are plotted with filled pentagons. Field very late M dwarfs of known 
distances are plotted with open pentagons. The three empty pentagons with 
error bars, joined with 
a dotted line, denote the position of Gl~569~B (combined light) and each of the 
two components resolved with the Keck AO system.}

\end{document}